# DFT based investigation of bulk mechanical, thermophysical and optoelectronic properties of $PbTaSe_2$ topological semimetal

A.S.M. Muhasin Reza, S.H. Naqib*
Department of Physics, University of Rajshahi, Rajshahi 6205, Bangladesh
*Corresponding author; email: salehnaquib@yahoo.com

## Abstract

$PbTaSe_2$ is a non-centrosymmetric topological semimetal. In this work we have explored the structural, elastic, mechanical, bonding, electronic, acoustic, thermal, and optical properties of $PbTaSe_2$. The electronic bond structure calculations confirm semi-metallic character. Fermi surface topology shows both electron and hole sheets. The single crystal elastic constants reveal that $PbTaSe_2$ is elastically stable. The compound is soft, brittle, and highly machinable at the same time. It also possesses very high level of dry lubricity. Various anisotropy indicators suggest that $PbTaSe_2$ is elastically anisotropic with layered character. The phonon dynamics has been investigated. Phonon dispersion plot shows that the compound is dynamically stable with a clear frequency gap between the acoustic and optical branches. The Debye temperature, phonon thermal conductivity, and melting temperature of $PbTaSe_2$ is low. The compound has medium Grüneisen parameter. The bonding character is mainly dominated by ionic bonding with some metallic contribution. The optical parameters have been studied in detail. The optical spectra reveal metallic features. The compound reflects visible light very efficiently (reflectance above 60%). It is also an efficient absorber of the ultraviolet light. The compound exhibits significant optical anisotropy with respect to the polarization directions of the incident electric field.

## 1. Introduction

Condensed matter physics is an important branch of material physics. Broadly speaking, there are three types of material based on their electronic structure based on the band theory of solids; insulators, semiconductors and metals. Materials are also classified in various phases based on symmetry states and symmetry braking states [1–6]. In topological materials specific states of electrons are topologically protected and are free from environmental perturbation. The compound $PbTaSe_2$ is a topological semimetal which becomes a non-centrosymmetric superconductor at 3.8 K. Structurally, the compound has Pb layers which are separated by the $TaSe_2$ layers [7,8]. In this structure the Pb atoms made triangular lattice and is sandwiched between the hexagonal $TaSe_2$ layers in the ground state. The compound $PbTaSe_2$ undergoes a number of structural phase transitions with different crystal symmetries at high pressures. The angle-resolved photoemission spectroscopy (ARPES) and first-principles calculations, showed the bulk nodal-line band structure and fully spin-polarized topological surface states in $PbTaSe_2$ [7–9].

The electronic properties of topological semimetals like the gapless band structure can be used in photo detectors [10–12]. Fermi-arcs for spintronics and qubits [13-16] for nanoscale device applications in nanostructures. Previously, the electronic band structure, energy density of states and superconducting state of $PbTaSe_2$ have been studied in detail [7,8,10,15,17,18]. On the other hand, investigation of various bulk physical properties, such as, elastic constants, mechanical characteristics, anisotropy, bonding features, optical properties, and thermal parameters in the ground state structure have not been explored theoretically yet. There is significant research gap existing as far as ground state bulk physical properties of $PbTaSe_2$ are concerned. We wish to fill these research gaps in this study.

The rest of the paper has been arranged as follows. The computational methodology is described in Section 2. The results of the calculations are presented and discussed in Section 3. Section 4 consists of the conclusions of this work.

## 2. Computational scheme

In this work all the calculations are carried out using the plane wave pseudopotential density functional theory (DFT) as implemented by the CAmbridge Serial Total Energy Package (CASTEP) [19–21]. The exchange-correlation terms are incorporated in the total energy by using the generalized gradient approximation (GGA) [22] with the functional developed by Perdew, Burke, and Ernzerhof (PBE-sol) [23] and the local density approximation (LDA). The ground state of a crystalline solid material is found by solving the Khon-Sham equation [20]. For reliable results, selection of pseudopotential is important. The pseudopotential gives the residual attractive interaction between an electron and an ion after taking into account the effective repulsion that arises from the exclusion principle demanding that valence states are orthogonal to core electronic states. The on-the-fly generated (OTFG) ultrasoft pseudopotential have been used in the calculations [24] .

The Broyden-Fletcher-Goldfarb-Shanno (BFGS) optimization method has been adopted to find out the ground state crystal structure. We have also used the density mixing [25]. The following valence electron orbitals are considered for Pb, Ta, Se atoms, respectively: Pb [$6s^2 6p^2$], Ta[$5d^3 6s^2$], Se[$4s^2 4p^4$]. The G-centered k-points have been considered in the reciprocal space (Brillouin zone). A k-points mesh of size 19×19×6 in the Monkhorst-Pack grid scheme [26] has been used for the sampling of the first Brilloun zone (BZ) of the hexagonal unit cell for $PbTaSe_2$. A plane wave basis with a cut off energy of 500 eV is used to expand the eigenfunctions of the valence and nearly valence electrons. Geometry optimization has been performed using a self-consistent convergence limit of $10^{-6}$ eV/atom for the energy, 0.02 eV/Å for the maximum force, 0.05 GPa for the maximum stress and $10^{-3}$ Å for maximum atomic displacement.

The optical properties of $PbTaSe_2$ have been evaluated using the ground state electronic band structure. The single crystal elastic constants have been calculated using the stress-strain method contained in the CASTEP. Thermal parameters have been studied with the help of elastic constants and moduli. The phonon dynamical calculations are carried out using the linear response theory. We have not included the spin-orbit coupling (SOC) in the band structure calculations. This is because the focus of this work is the bulk physical properties of $PbTaSe_2$ which are not greatly affected by the SOC [3,4,6,27]. Inclusion of the SOC affects mainly the surface electronic states. The main effect of SOC is the splitting of surface sensitive electronic states of $PbTaSe_2$ by ~ 0.10 to 0.20 eV with an amplification of the topological semimetal characteristics of the compound. Such splitting does not affect the bulk physical behavior significantly.

The chemical bonding natures of $PbTaSe_2$ have been explored via the Mulliken population analysis (MPA) and the Hirshfeld population analysis (HPA).

## 3. Results and analysis

### 3.1. Structural properties

The $P\bar{6}m_2$ structure of $PbTaSe_2$ is called the α-phase. The schematic hexagonal [space group $P\bar{6}m_2$ (No 187)] crystal structure of $PbTaSe_2$ is shown in Figure 1. The unit cell consists of 4 atoms in which there is one Pb atom, one Ta atom and two Se atoms. The atomic positions and lattice parameters of the crystal are fully optimized starting with the experimental values found in earlier studies [28]. The calculated lattice constants a (= b) and c along with experimental and other theoretical values are given in Table 1. It is observed that the present values are very close to the experimental ones [7,28–30]. Since optimization of the crystal geometry is one of the most crucial part in any ab-initio investigation, excellent agreement between the computed and experimental lattice constants imply that the results obtained in this study are reliable [31].

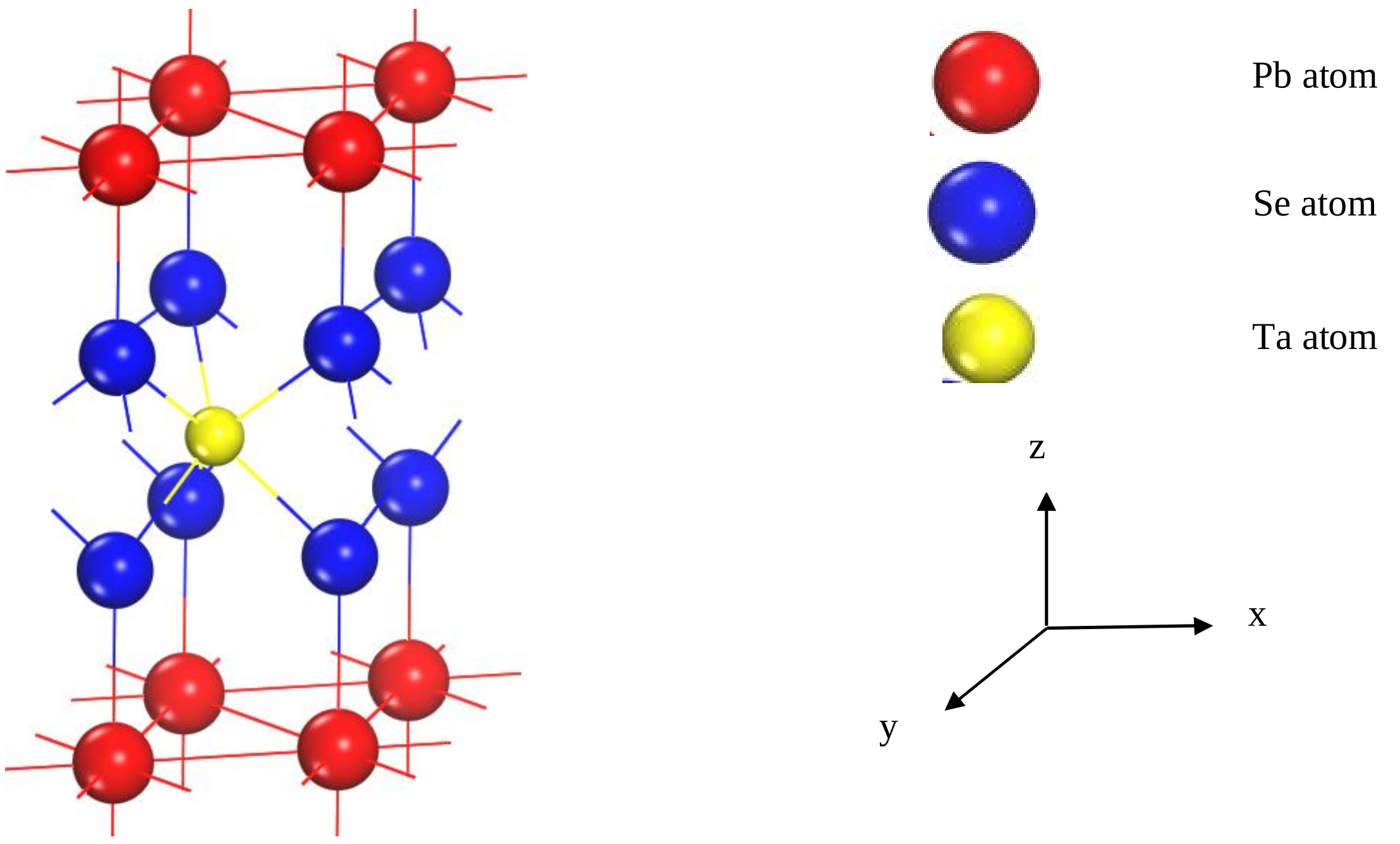

**Figure 1:** Schematic crystal structure of $PbTaSe_2$. The crystallographic directions are shown.

The lattice can be viewed as a Pb atomic layer is interspaced with two adjacent $TaSe_2$ layers. The Pb-Se bonds are comparatively weak compared to the Ta-Se bonds. The structure of $PbTaSe_2$ is highly sensitive to the external pressure. There are three other different structures of $PbTaSe_2$ which are found at different elevated pressures and temperatures. These are the $P6_3mc$ (β-phase), P6/mmm (γ-phase) and Pmmm (δ-phase). In this work, we are concerned with the $P\bar{6}m_2$ structure of $PbTaSe_2$ which is stable under normal pressure and temperature.

**Table 1:** Calculated and lattice constants a = b and c, c/a ratio, and equilibrium cell volume of hexagonal $PbTaSe_2$.

| Compound | a (Å) | c (Å) | c/a | Volume $V_0$ (Å$^3$) | Ref. |
|---|---|---|---|---|---|
| $PbTaSe_2$ | 3.40 | 9.37 | 2.75 | -- | [28] Exp. |
| | 3.39 | 9.29 | 2.74 | -- | [28] Theo. |
| | 3.44 | 9.35 | 2.71 | -- | [29,31] exp. |
| | 3.41 | 9.38 | 2.74 | -- | [29] exp. |
| | 3.45 | 9.35 | 2.71 | -- | [30] exp. |
| | 3.43 | 9.38 | 2.73 | -- | [30] exp. |
| | 3.41 | 9.34 | 2.73 | 94.5 | This work |

## 3.2. Elastic properties

### *3.2.1. The stiffness constants*

All the mechanical properties of crystalline solids are determined by the elastic constants ($C_{ij}$). The possible applications are also limited by the elastic behavior. The elastic constants ($C_{ij}$) are connected with the bonding characteristics of materials. The elastic constants also determine the mechanical stability of a solid. The bulk elastic behavior is understood from the polycrystalline elastic moduli. The compound under study is hexagonal and, therefore, it has five independent elastic constants: $C_{11}$, $C_{12}$, $C_{13}$, $C_{33}$, and $C_{44}$. The elastic constant $C_{66}$ is not independent and can be expressed as $C_{66} = \frac{C_{11} - C_{12}}{2}$. The mechanical stability conditions of a hexagonal crystal system can be expressed as follows [32,33].

$$C_{11} > 0;\ C_{11} > C_{12};\ C_{44} > 0;\ (C_{11}+C_{12})C_{33}-2(C_{13})^2 > 0 \qquad (1)$$

Above conditions are satisfied by $PbTaSe_2$, and hence the compound is expected to be mechanically stable. The computed single crystal elastic constants are given in Table 2. There is no available data on the elastic constants of orthorhombic $PbTaSe_2$ in the $P\bar{6}m_2$ structure, to the best of our knowledge.

**Table 2:** The calculated elastic constants, $C_{ij}$ (GPa), of $PbTaSe_2$ in the ground state.

| Compound | $C_{11} = C_{22}$ | $C_{12}$ | $C_{13} = C_{23}$ | $C_{33}$ | $C_{44} = C_{55}$ | $C_{66}$ |
|---|---|---|---|---|---|---|
| $PbTaSe_2$ | 129.5 | 38.36 | 11.17 | 144.43 | 13.08 | 45.57 |

There three diagonal elastic constants $C_{11}$, $C_{22}$ and $C_{33}$ give the measure of the resistance of the crystal to the uniaxial stress in the a-, b- and c-axis, respectively. The non-diagonal elastic tensors are related to the shape deforming shearing stresses. Among these non-diagonal elastic constants, $C_{44}$ is closely linked to the hardness of the compound. Very low value of $C_{44}$ implies that $PbTaSe_2$ is a soft material. Different values of $C_{11}$ (= $C_{22}$) and $C_{33}$ is a reflection of unequal bonding strengths in $PbTaSe_2$ in the ab-plane and c-direction.

### *3.2.2. Elastic moduli and parameters*

The elastic moduli in the polycrystalline state can be calculated from the single crystal elastic constants using the well known standard formulae [34–41]. Some other useful bulk elastic indicators, e.g., the Pugh's ratio (k), Poisson's ratio (σ), and the machinability index ($\mu_M$) can be calculated from the elastic moduli as follows [39-41]:

$$k = \frac{B}{G} \quad (2)$$

$$\sigma = \frac{3B - 2G}{6B + 2G} \quad (3)$$

$$\mu_M = \frac{B}{C_{44}} \quad (4)$$

In Eqns. 2 – 4, B and G denote the bulk modulus and the modulus of rigidity, respectively. The macro hardness ($H_{macro}$) and the micro hardness ($H_{micro}$) parameters of $PbTaSe_2$ are also calculated using the following formulae [39,42,43].

$$H_{macro} = 2[(\frac{G}{B})^2 G]^{0.585} - 3 \quad (5)$$

$$H_{micro} = \frac{(1-2s)Y}{6(1+s)} \quad (6)$$

In Eqn. 6, Y is the Young's modulus of the solid. The Cauchy pressure, $C_p$, is another useful elastic parameter closely related to the bonding character and brittleness/ductility of solids. For hexagonal crystals, $C_p = (C_{12} - C_{44})$. The computed elastic moduli and parameters are shown in Table 3 below.

**Table 3:** The computed bulk modulus B (GPa), shear modulus G (GPa), Young's modulus Y (GPa), machinability index $\mu_M$, macro hardness $H_{macro}$ (GPa), micro hardness, $H_{micro}$ (GPa), Pugh's ratio k, Poisson's ratio σ and the Cauchy pressure $C_p$ (GPa) of $PbTaSe_2$.

| Compound | B | G | Y | $\mu_m$ | $H_{macro}$ | $H_{micro}$ | k | σ | $C_p$ |
|---|---|---|---|---|---|---|---|---|---|
| $PbTaSe_2$ | 58.27 | 30.43 | 77.75 | 4.45 | 4.51 | 3.49 | 1.91 | 0.27 | 25.28 |

The machinability index measures the ease with which a solid can be cut into desired shapes. It also measures the level of plasticity and dry lubricity of a compound [44–47]. The elastic moduli B, G and Y quantify the resistance of the bulk material to volume, shape and length changing deformations. It is seen that the elastic moduli of $PbTaSe_2$ are low. The machinability index, on the other hand is very high. This implies that the compound under consideration is highly machinable and has significant dry lubricity. Both the macro and micro hardnesses are low suggesting that the overall bonding strength in $PbTaSe_2$ is weak. The Pugh's ratio can distinguish between the mechanical failure modes of materials. Solids having a Pugh's ratio greater than 1.75 are ductile in nature. If the Pugh's ratio is below this value, then the material is predicted to be ductile [42]. The obtained value of Pugh's ratio of $PbTaSe_2$ is 1.91 which is greater than 1.75. Therefore, $PbTaSe_2$ should show ductility. The nature of bonding can be gauged from the Poisson's ratio. The central force interactions dominate in solids when the value of Poisson's ratio lies between 0.25 to 0.50 [48]. For $PbTaSe_2$, the value of Poisson's ratio is 0.27 meaning that the chemical bondings have central force nature. Furthermore, for a purely covalent crystal the value of Poisson's ratio is around 0.10 and for a crystal with metallic bonding, the value of Poisson’s ratio is around 0.33 [49]. Therefore, we expect notable contribution of metallic bonding in $PbTaSe_2$. The value of the Poisson’s ratio can also be used to determine brittleness/ductility. If σ is less than the critical value 0.26, then a material is predicted to be brittle; it will be ductile otherwise. Poisson’s ratio is 0.27 for $PbTaSe_2$ indicating the ductile nature. The Cauchy pressure is another tool used to separate the materials into brittle or ductile group where the critical value of $C_p$ is zero [50]. The negative value of $C_p$ indicates that the solid is brittle in nature. The obtained value of $C_p$ of $PbTaSe_2$ is positive signifying that our material is ductile in nature. According to the Pettifor’s rule [51] materials with large positive Cauchy pressure have significant metallic bonds. On the contrary, negative Cauchy pressure suggests that angular bonds are dominating which makes a compound brittle. The results presented in Table 3 are novel and therefore, no comparison with any other reported values can be made.

### *3.2.3. Elastic anisotropy*

Elastic anisotropy controls many important mechanical properties of solids related to their practical applications [14]. The elastic anisotropy of $PbTaSe_2$ has been studied by means of different anisotropy indices.

The shear anisotropy factors measure bonding strength among atoms in different planes. There are three different shear anisotropy factors for hexagonal crystal. These factors can be calculated from the following equations [52]:

$$A_1 = \frac{(C_{11} + C_{12} + 2C_{33} - 4C_{13})}{6C_{44}} \qquad (7)$$

for {100} shear planes between <011> and <010> directions.

$$A_2 = \frac{2C_{44}}{C_{11} - C_{12}} \qquad (8)$$

for {010} shear planes between < 101> and <001> directions and

$$A_3 = \frac{C_{11} + C_{12} + 2C_{33} - 4C_{13}}{3(C_{11} - C_{12})} \qquad (9)$$

for {001} shear planes between <110> and <010> directions.

The shear anisotropy factors $A_i$ (i = 1, 2, 3) have the unit value for isotropic crystals. Departure from unity quantifies the level of anisotropy in shape changing deformation.

The directional bulk modulus along a-direction and c-direction can be calculated by using the following relations [53]:

$$B_a = \alpha\frac{dP}{da} = \frac{L}{(2 + a)} \qquad (10)$$

$$B_c = \alpha\frac{dP}{dc} = \frac{B_a}{a} \qquad (11)$$

$$\text{where, } \Lambda = 2(C_{11}+C_{12}) + 4C_{13}\alpha + C_{33}\alpha^2 \qquad (12)$$

$$\text{and } \alpha = \frac{(C_{11} + C_{12} - 2C_{13})}{(C_{33} + C_{13})} \qquad (13)$$

In addition to the above anisotropy indicators, the ratio of the two linear compressibility coefficients along a- and c- axis of hexagonal crystals, $K_c/K_a$, is another useful parameter to understand the in-plane and out-of-plane anisotropy in the bonding strengths. The relevant expressions are [52].

$$K_c = (C_{11}+C_{12} - C_{13}) \qquad (14)$$

$$K_a = (C_{33} - C_{13}) \qquad (15)$$

$$K_c/K_a = \frac{(C_{11}+C_{12}-2C_{13})}{(C_{33}-C_{13})} \qquad (16)$$

All these anisotropy factors are evaluated and listed in Table 4. $B_a = B_c$ for isotropic crystals and and $K_c/K_a = 1$, for the same.

**Table 4:** Shear anisotropy factors ($A_1$, $A_2$ and $A_3$), directional bulk moduli ($B_a$, $B_c$ in GPa) and ratio of linear compressibility coefficients ($K_c/K_a$) of $PbTaSe_2$.

| Compound | $A_1$ | $A_2$ | $A_3$ | $B_a$ | $B_c$ | $K_c/K_a$ |
|---|---|---|---|---|---|---|
| $PbTaSe_2$ | 5.24 | 0.28 | 1.50 | 171.65 | 183.58 | 1.09 |

Table 4 reveals that the shear anisotropy is significant in $PbTaSe_2$. The anisotropies in the bulk moduli and compressibilities are moderate.

### 3.4. Electronic Properties

#### *3.4.1. Electronic band structure*

Electronic band structure is one of the most important features of solids which control all the electronic and optical properties. It also gives information regarding bonding in a crystal and electronic stability of a system. The bulk electronic band structure as a function of energy ($E-E_F$) along the high symmetry directions in the reciprocal space is calculated in the ground state and is shown in Figure 2. The Fermi level ($E_F$) is indicated by the horizontal broken line which was set at zero eV. The bulk electronic band structure shows several interesting behavior. The compound is semimetallic due to weak crossing of the energy bands of the Fermi level around the G-point. The conduction band is hole-like in this case. There is clear signature of semimetallic topological behavior close to the H-point along the A-H line where two nearly linear band segments touch each other exactly on the Fermi level.

The bands close to the Fermi level originate from the Pb-6p and Pb-6s and Se-4p and Se-4s electronic orbitals. Contribution from the electrons of the Ta atom is minute. There is significant sp-hybridization.

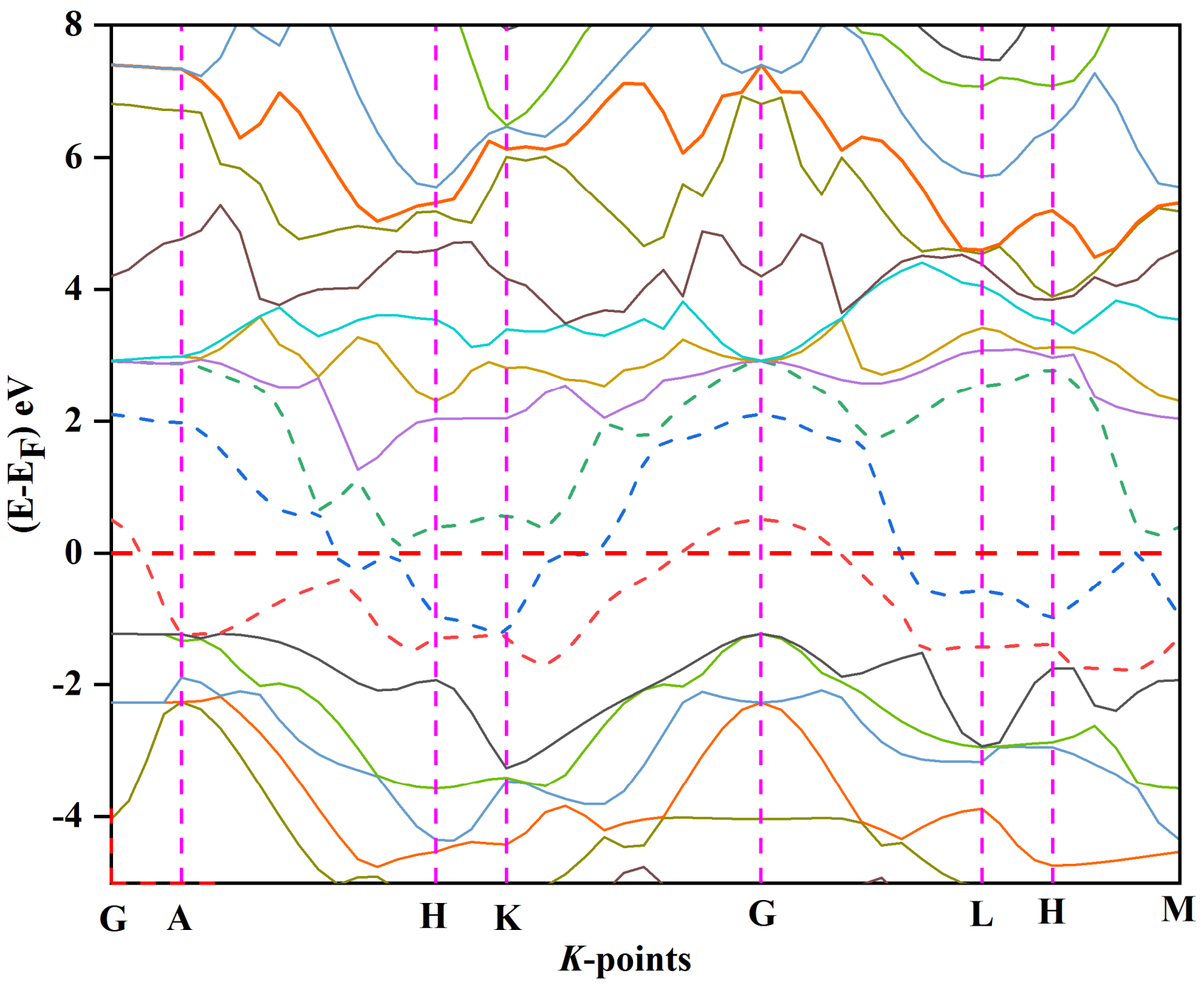

**Figure 2:** Electronic band structure of $PbTaSe_2$ in the ground state. The dashed horizontal line marks the Fermi energy (set to 0 eV).

### *3.4.2. Electronic energy density of states (DOS)*

The total and partial densities of states (PDOS and TDOS, respectively) of $PbTaSe_2$ are calculated from the ground state electronic band structure results. The PDOS and TDOS plots are given in Fig. 3. The vertical broken line indicates the Fermi level. The non-zero value of TDOS at the Fermi level confirms that $PbTaSe_2$ will exhibit metallic electrical conductivity. To get the contribution of each atom to the TDOS of $PbTaSe_2$, we have shown the PDOS of Pb, Ta and Se atoms. The TDOS value at the Fermi level is 1.86 states/eV-unit cell. The large peaks in the TDOS in the valence band centered at -3.5 eV and at 3.0 eV in the conduction band are principally responsible for charge transport and optoelectronic properties of $PbTaSe_2$. These two peaks are due to the Pb-6s, Pb-6p, Se-4s and Se-4p electronic states. The overall contribution of the electronic states of the Ta atom in the energy range shown in Fig. 3 is small. The Fermi level is located to the left of a pseudogap at 0.51 eV in the bonding region. This suggests that $PbTaSe_2$ has high electronic and structural stability.

The degree of electronic correlation in $PbTaSe_2$ can be estimated from the TDOS at the Fermi level, $N(E_F)$. The repulsive Coulomb pseudopotential, $V_c$, is a measure of the electronic correlation which is related to $N(E_F)$ as follows [54]:

$$V_c = 0.26N(E_F)/[1+N(E_F)] \qquad (17)$$

The calculated value of $V_c$ turns out to be 0.15. This shows that electronic correlation is not weak in $PbTaSe_2$.

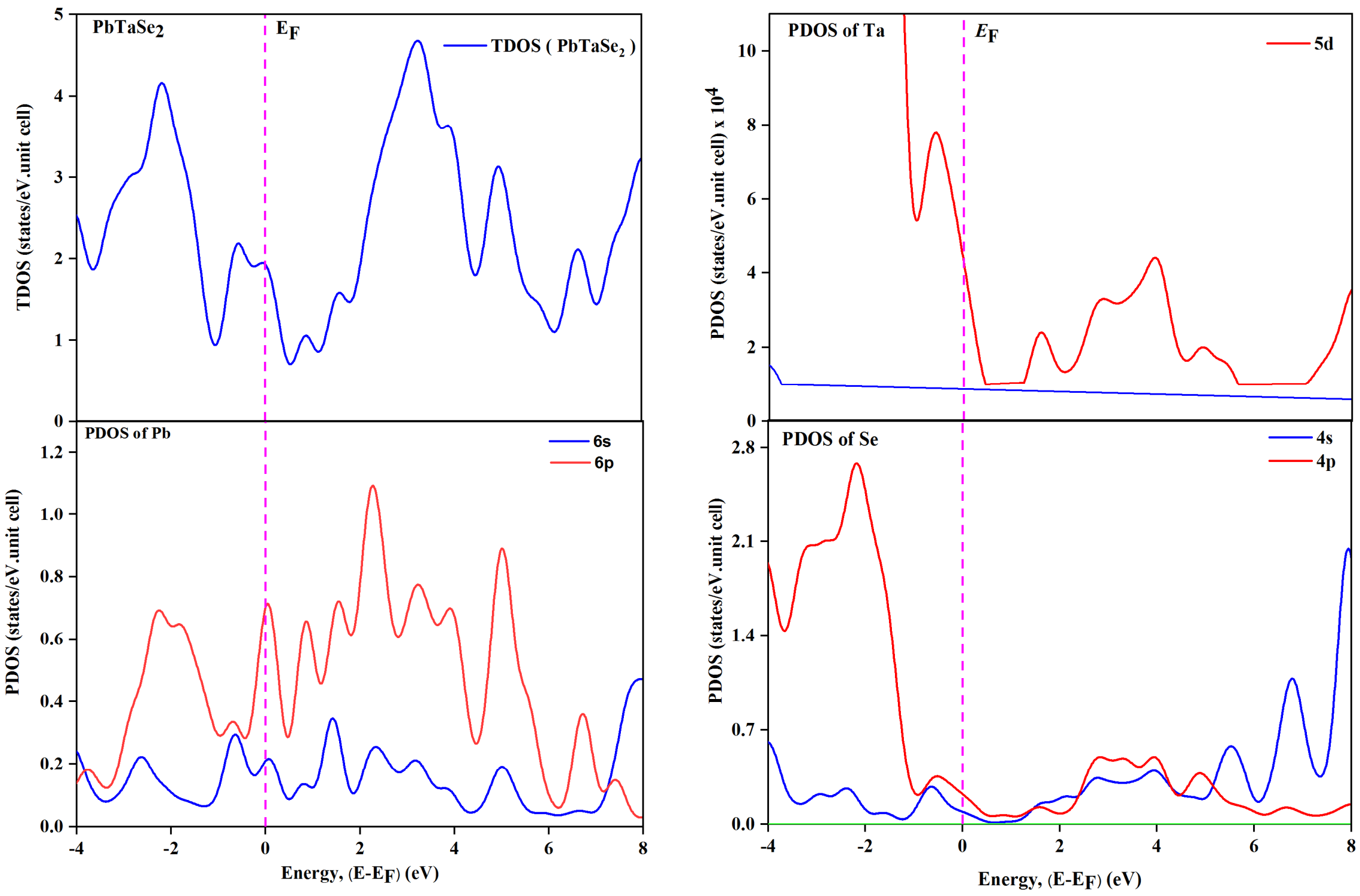

**Figure 3:** Total and partial electronic densities of states (TDOS and PDOS) of $PbTaSe_2$ in the ground state.

### *3.4.3. Electronic charge density distribution*

The charge distribution around the atoms inside the crystal is very important to understand the bonding nature. In this section we have studied the electronic charge density distribution in different crystal planes of $PbTaSe_2$. The electronic charge density distribution of $PbTaSe_2$ in the (111) plane and (001) planes are shown in Fig. 4**.** The color scales on the right hand side of the panels represent the total electron density. The blue color indicates the high charge (electron) density and the red color indicates the low charge (electron) density. The charge density maps show that the atoms in the (111) planes are mainly bound by metallic bonding. The same is true for the atoms in the (001) plane with slight tendency of covalent bonding. This tendency is understood from little deviation of the shapes of charge density distribution around Ta and Se atoms. From the charge density maps of $PbTaSe_2$ in both the planes, we can see that Ta and Se atoms have high electron density compared to the Pb atoms. The low charge concentration for the Pb atoms implies that the uniform background charges (the red region) probably come primarily from the Pb electrons in the conduction band.

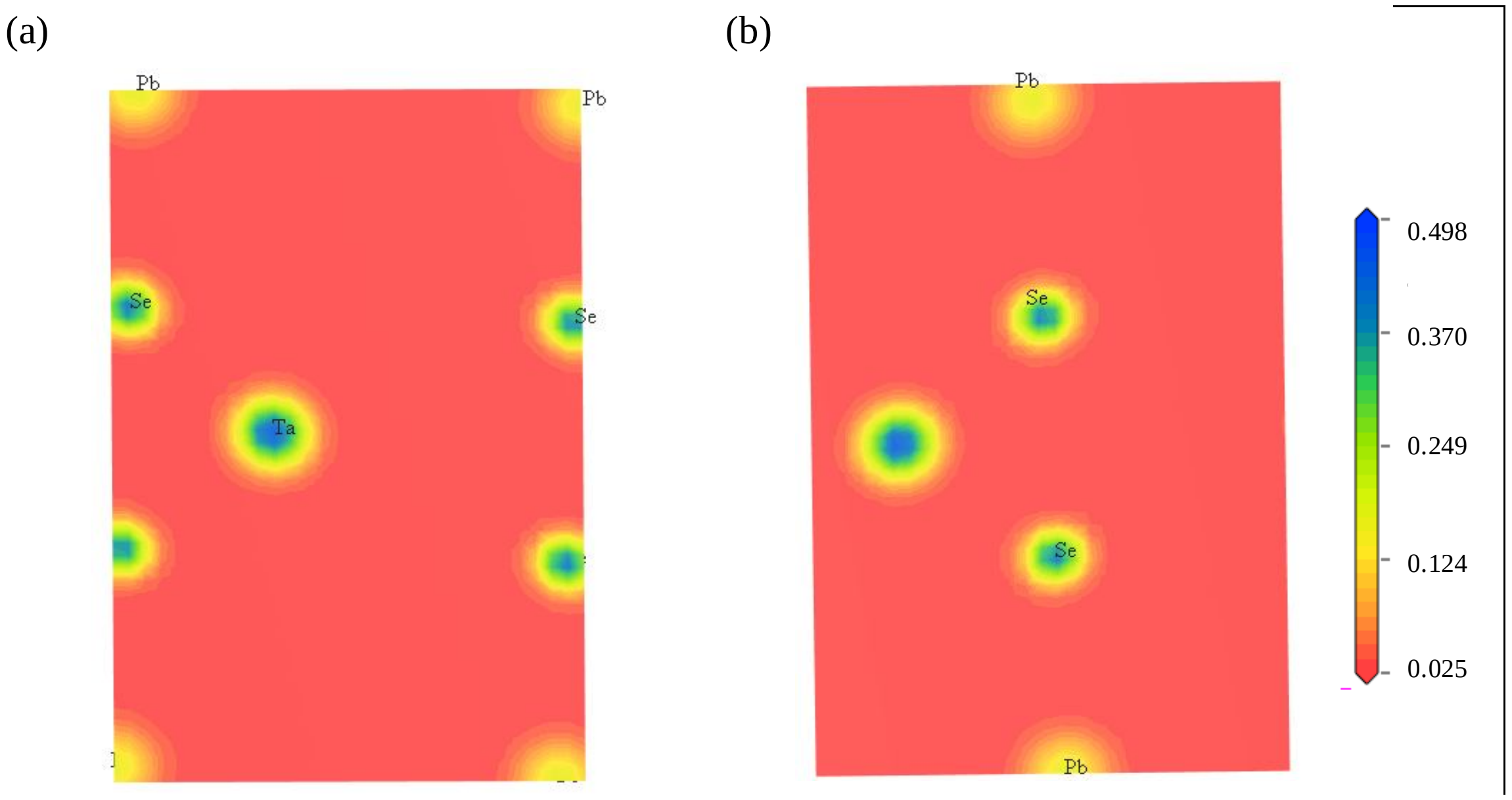

**Figure 4:** The electronic charge density distribution maps for $PbTaSe_2$ in the (a) (111) and (b) (001) planes. The color scale on the left quantifies the amount of charge (in unit of electronic charge).

### *3.4.4. Fermi surface*

The Fermi surface of $PbTaSe_2$ is shown in Fig. 4. From the band structure of $PbTaSe_2$, we have seen that three bands cross the Fermi level. These three bands contribute to the three sections of the Fermi surface. Weak crossing at around the H point in the BZ gives rise to the six small segments of electronic Fermi surface with hexagonal symmetry. The Fermi sheet enclosing the central ellipsoid has hole-like character. The central ellipsoid is electronic. This central part is expected to control the charge transport and optical properties of $PbTaSe_2$. Deviation from spherical shape of the Fermi sheets indicates that the electronic properties of $PbTaSe_2$ are anisotropic.

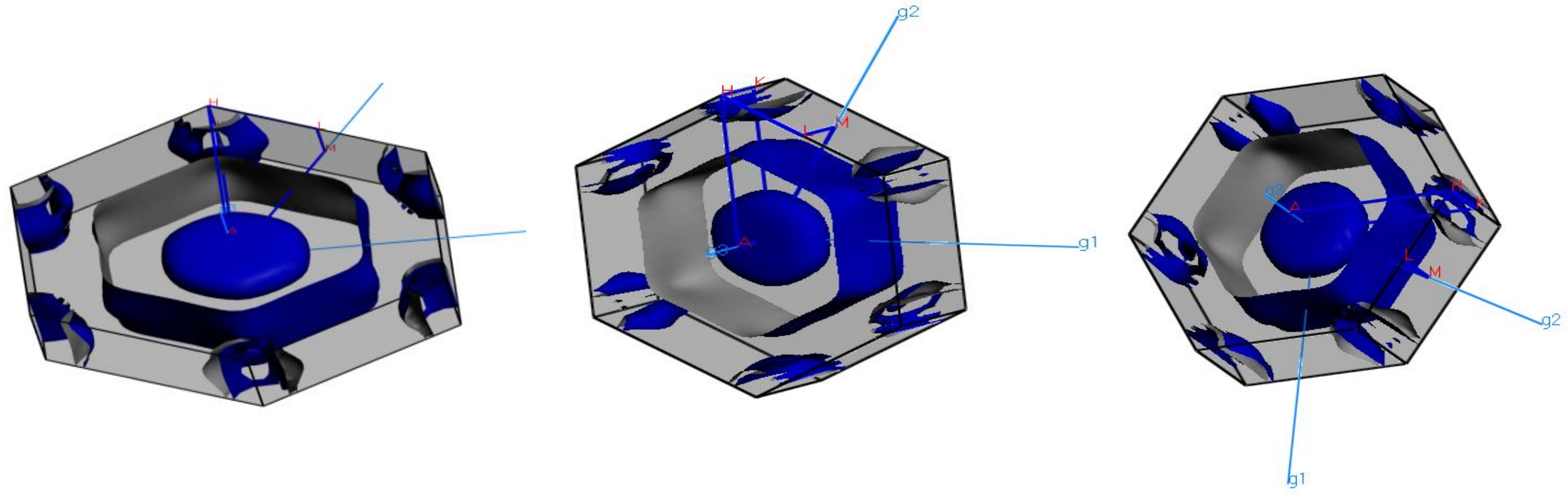

**Figure 5:** Fermi Surface of $PbTaSe_2$ seen from different angles.

### 3.5. Phonon and acoustic properties

#### *3.5.1. Phonon dispersion curves*

Phonons are the quanta of lattice vibrations. Large number of electrical, thermal and lattice dynamical properties of crystalline solids are dependent on the phonon spectrum. The phonon spectrum is determined by the crystal symmetry, stiffness constants and mass of the constituent atoms [55–57]. We have calculated the phonon dispersion curves and the phonon density of states (PHDOS) of the $PbTaSe_2$ compound along the high symmetry directions of the BZ. The calculated curves are shown in Fig. 6.

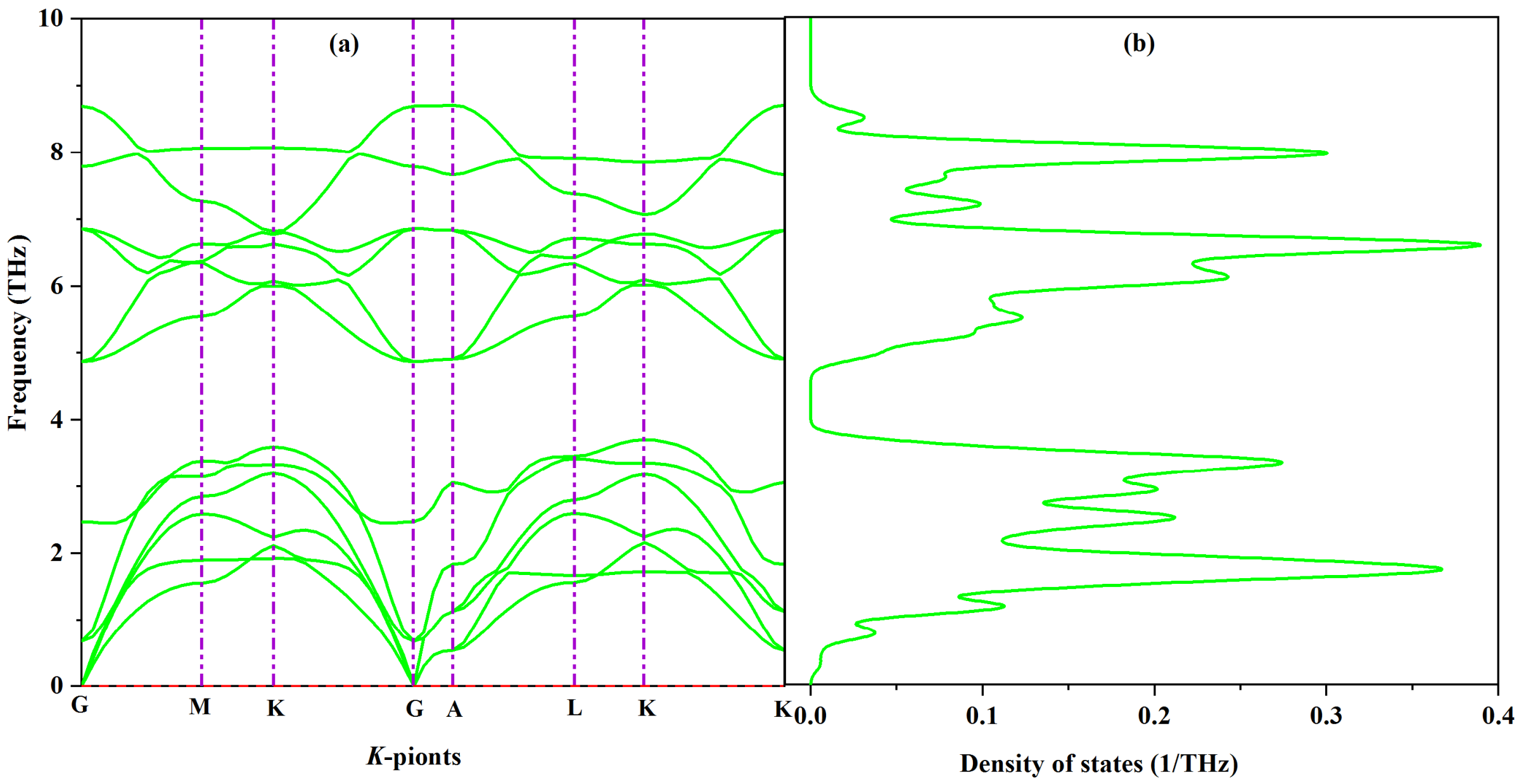

**Figure 6:** Calculated (a) phonon dispersion spectra and the (b) PHDOS for $PbTaSe_2$ compound at zero pressure.

As can be seen from Fig. 6a, no negative energy phonon branch exists in the dispersion curves. This is a clear indication of the dynamical stability of the compound under study. The acoustic and the optical branches are separated by a frequency gap. The high PHDOS regions for the acoustic branches at low phonon frequencies should contribute significantly in the thermal transport. The highest energy phonon dispersion branches are due to the vibration of light Se atoms in the crystal.

#### *3.5.2. Acoustic properties*

The sound velocity through a material is very important in relation to the thermal and electrical conduction. The average sound velocity in solids, $v_m$, is related to the shear modulus (G) and bulk modulus (B). The $v_m$ is given by the harmonic mean of the average longitudinal and transverse sound velocities, $v_l$ and $v_t$, respectively.

$$v_m = [\frac{1}{3}(\frac{1}{V_l^3} + \frac{2}{V_t^3})]^{-1/3} \quad (18)$$

The velocities $v_l$ and $v_t$ are calculated from the following equations [40]:

$$v_l = [\frac{3B + 4G}{3r}]^{1/2} \quad (19)$$

$$\text{and } v_t = [\frac{G}{r}]^{1/2} \quad (20)$$

Table 7 exhibits the calculated crystal density of and the acoustic velocities in $PbTaSe_2$.

**Table 7:** Density ρ ($kg/m^3$), transverse velocity $v_t$ ($ms^{-1}$), longitudinal velocity $v_l$ ($ms^{-1}$) and average elastic wave velocity $v_m$ ($ms^{-1}$) of $PbTaSe_2$.

| Compound | ρ | $v_l$ | $v_t$ | $v_m$ |
|---|---|---|---|---|
| $PbTaSe_2$ | 9593.9 | 3209.7 | 1781.3 | 1881.7 |

## 3.6. Thermal properties

### *3.6.1. Debye temperature*

The Debye temperature is one of the most prominent thermophysical parameter of a material. It is connected to the phonon thermal conductivity, heat capacity, melting temperature, superconducting transition temperature, and electrical conductivity of solids. There are several methods for calculating the Debye temperature, $\theta_D$. Among those, the one proposed by Anderson [58] is simple and results in reliable estimation of Debye temperature solids belonging to different crystal symmetries. The relevant expression is given below:

$$\theta_D = \frac{h}{k_B}[(\frac{3n}{4p})\frac{N_A r}{M}]^{1/3} v_m \quad (21)$$

In the above equation, h is the Planck's constant and $k_B$ is the Boltzmann constant, n refers to the number of atoms in a molecule, $N_A$ is the Avogadro's number, ρ is the mass density, M is the molecular mass and $v_m$ is the average velocity of sound in the crystalline solid. The calculated Debye temperature is given in Table 8. The Debye temperature of $PbTaSe_2$ is low, 216.53 K. Thus the phonon thermal conductivity of this compound is expected to be low as well. This low value also points towards weak bonding among the atoms of $PbTaSe_2$.

### *3.6.2. The melting temperature*

Information on the melting temperature ($T_m$) is necessary for materials to be used at elevated temperatures. The melting temperature is related to the bonding strength and cohesive energy of the crystal. These parameters also determine the elastic constants. Fine et al. [59] developed a formula for calculating the melting temperature of hexagonal crystal using the elastic constants as follows:

$$T_m = 354+1.5(2C_{11}+C_{33}) \quad (22)$$

The calculated melting temperature of $PbTaSe_2$ is listed in Table 8. The relatively low melting temperature (959.16 K) of $PbTaSe_2$ is consistent with its low Debye temperature, hardness and various elastic moduli.

### *3.6.3. Thermal conductivity*

Thermal conductivity is an important thermal transport coefficient that gives the measure of the efficiency of heat transfer by a material. This parameter is temperature dependent. The minimum thermal conductivity ($K_{min}$) is the limiting value of the thermal conductivity at high temperature when the phonon contribution to the thermal conductivity ($K_{ph}$) reaches its minimum value and becomes independent of temperature. Based on the Debye model, Clarke derived the formula to calculate the minimum thermal conductivity ($K_{min}^{Clarke}$) using the following equation [60]:

$$K_{min}^{Clarke} = k_B v_m \left[\frac{nrN_A}{M}\right]^{2/3} \qquad (23)$$

The minimum thermal conductivity can also be estimated employing the Cahill formalism where the phonon spectrum has been considered within the Einstein model. The Cahill formula [61] for minimum thermal conductivity is given below:

$$K_{min}^{Cahill} = \frac{k_B}{2.48} n^{\frac{2}{3}} (v_l + 2v_t) \qquad (24)$$

The calculated minimum thermal conductivities of $PbTaSe_2$ are tabulated below (Table 8). The minimum thermal conductivity of $PbTaSe_2$ is very low. Semimetallic character of $PbTaSe_2$ implies that the electronic contribution to the thermal conductivity at high temperature should also be low. Thus the overall low thermal conductivity of $PbTaSe_2$ can make it a useful thermal barrier material.

The temperature dependent phonon thermal conductivity of $PbTaSe_2$ can be estimated using the scheme developed by Slack [62]. The Slack formula for temperature dependent phonon thermal conductivity is as follows:

$$K_{ph}(T) = A\left[\frac{M_{av}\theta_D^3\delta}{\gamma^2 N^{2/3} T}\right] \qquad (25)$$

In Eqn. 25, $M_{av}$ is the average atomic mass (Kg/mol) in the molecule, $\delta$ denotes the cubic root of average atomic volume, N denotes the number of atoms present in unit cell, and $\gamma$ denotes the Grüneisen constant calculated using the poison's ratio ($\sigma$). *A* is a parameter (in W-mol/Kg/m$^2$/K$^3$) depending on $\gamma$ which is calculated from [63]:

$$A(\gamma) = \frac{5.720 \times 10^7 \times 0.849}{2 \times \left[1 - \frac{0.514}{\gamma} + \frac{0.224}{\gamma^2}\right]} \qquad (26)$$

The Grüneisen parameter gives the measure of anharmonicity in a solid. The Grüneisen parameter $\gamma$ is an important quantity in the thermodynamics and lattice dynamics because it is related with bulk modulus, heat capacity at constant volume, thermal expansion coefficient, lattice thermal conduction and volume of the solid. The Grüneisen parameter can be evaluated with the Poisson's ratio as follows [64]:

$$\gamma = \frac{3[1+\sigma]}{2[2-3\sigma]} \qquad (27)$$

For crystalline materials the values of $\gamma$ is usually found in the range of 0.80 - 3.50 [65-67]. The calculated value of $\gamma$ of $PbTaSe_2$ is 2.24 which is well within the established range. The value is in the medium range which implies the medium level of anharmonic effects in $PbTaSe_2$. All the thermophysical parameters calculated in this section are presented in Table 8 below.

**Table 8:** The Debye temperature $\theta_D$, minimum thermal conductivity $K_{min,}$ Grüneisen parameter $\gamma$, lattice thermal conductivity $K_{ph}$ at 300 K and the melting temperature $T_m$ of $PbTaSe_2$ compound.

| Compound | $\theta_D$ (K) | $K_{min}$ (W/m-K) | | $\gamma$ | $K_{ph}$ (W/m-K) | $T_m$ (K) |
|---|---|---|---|---|---|---|
| | | $K_{min}^{Clark}$ | $K_{min}^{Cahill}$ | | | |
| $PbTaSe_2$ | 216.5 | 0.414 | 0.468 | 2.24 | 11.36 | 959.16 |

### 3.7. Bond population analysis

The Mulliken bond populations are calculated to understand the bonding nature (ionic, covalent and metallic) of $PbTaSe_2$. We have also performed the Hirshfeld population analysis (HPA). The calculated values of atomic populations and relevant parameters are presented in Table 9. It is seen from Table 9 that in $PbTaSe_2$, electrons are transferred from Pb and Ta to Se. It is an indication of ionic bonds. The electron transfer can be attributed to the difference in the electron affinities of Se, Ta, and Pb. On the other hand, non-zero effective valence charge implies that there is some covalent contributions as well. The effective valences in the Mulliken population analysis (MPA) and the HPA are different. This is actually expected since MPA depends on the basis sets used to approximate the wave functions of the orbitals while the HPA is independent of the basis sets.

**Table 9:** Charge Spilling parameter (%), orbital charge (electron), atomic Milliken charge (electron), effective valance (Mulliken & Hirshfeld charge) (electron) of $PbTaSe_2$.

| Charge Spilling (%) | Atom | Mulliken atomic population | | | | | Mulliken charge | Formal ionic charge | Effective valence (Mulliken) | Effective valence (Hirshfeld) |
|---|---|---|---|---|---|---|---|---|---|---|
| | | s | p | d | F | total | | | | |
| 3.66 | Pb | 3.33 | 8.21 | 10.0 | 0.0 | 21.54 | 0.46 | +2 | 1.54 | 1.92 |
| | Ta | 2.00 | 6.00 | 0.0 | 14 | 22.00 | 5.00 | +4 | -1.00 | 0.92 |
| | $Se_1$ | 2.73 | 5.70 | 10 | 0 | 18.44 | -2.44 | -2 | -0.44 | 1.92 |
| | $Se_1$ | 2.73 | 5.70 | 10 | 0 | 18.44 | -2.44 | -2 | -0.44 | 1.92 |

### 3.8. Optical properties

We have studied the optical properties of $PbTaSe_2$ to explore the possible scopes of its use in the optical sector. In this section we have calculated the optical properties such as absorption coefficient, dielectric constant, photoconductivity, refractive index, reflectivity and loss function in the photon energy range up to 20 eV with two different polarization directions of [100] and [001]. The methodology for optical calculations is detailed in Refs. [68,69]. A Gaussian smearing of 0.5 eV, a Drude energy 0.05 eV and an unscreened plasma energy of 10 eV were used to calculate the optical parameters as a function of incident photon energy. All the computed optical parameters are shown in Fig. 7 below.

The real part of dielectric function, $\varepsilon_1(\omega)$, is shown in Fig.7a**.** The spectra of $\varepsilon_1$ start from negative value with a peak around ~3.5 and cross the zero line at around 17.5 eV. This is a typical metallic behavior. Fig. 7a also shows the imaginary part of the dielectric function, $\varepsilon_2(\omega)$. This is related to the photon absorption characteristics of $PbTaSe_2$. The peaks and the spectral weight in $\varepsilon_2(\omega)$ are controlled by the electronic energy density of states of the energy levels involved in the optical transition of electrons and the matrix elements of the transition between the two states involved. Sharp peaks are found at 4.95 eV for [100] and at 5.03 eV for [001] polarization directions of the incident electric field vector. For both polarizations $\varepsilon_2$ gradually decreases with increasing energy and finally goes to zero at ~ 18 eV. There is significant optical anisotropy in the dielectric function with respect to the polarization states of the electric field.

The real part of refractive index, $n(\omega)$, and the imaginary part, $\kappa(\omega)$, are shown in Fig. 7b. The low energy value of $n(\omega)$ was found to be high in the infrared and visible region. This real part measures the group velocity of electromagnetic wave in the solid. The imaginary part, also known as the extinction coefficient, measures the attenuation of light as it travels through the material. From Fig. 7b it is seen that visible light is highly attenuated by $PbTaSe_2$. Both the real and imaginary parts of the refractive index decrease monotonically at high energies in the ultraviolet (UV) region of the electromagnetic spectrum. The optical anisotropy is quite pronounced up to 10 eV.

The variation of absorption coefficient, $\alpha(\omega)$, as a function of photon energy is shown in Fig. 7c. Finite values of $\alpha(\omega)$ for both the polarizations at very low energy indicate the metallic state of $PbTaSe_2$. The absorption coefficient is quite high in the energy range 5 eV to 15 eV in the UV region. This suggests that $PbTaSe_2$ is a good absorber of ultraviolet radiation. There is significant optical anisotropy in absorption in the energy range 5 eV to 10 eV.

The photoconductivity is an important parameter for optoelectronic device applications. Optical conductivity as a function of photon energy is depicted in Fig.7 d. The low energy photoconductivity reaffirms the metallic character of $PbTaSe_2$. There is significant optical anisotropy in $\sigma(\omega)$.

The reflectivity, as a function of incident photon energy, is given in Fig. 7e. The reflectivity is high in the infrared and visible region. The reflectivity initially decreases in the near-infrared region then increases gradually and becomes almost non-selective in the energy range 5 eV to 15 eV. Thus $PbTaSe_2$ has high potential to be used as an efficient solar reflector to reduce solar heating. $R(\omega)$ decreases sharply at around 18 eV close to the plasma peak in the energy loss spectrum.

The calculated energy loss spectrum is shown in Fig. 7f. The energy loss function helps one to understand the screened plasma excitation made by swift charges moving inside the material. The loss function, $L(\omega)$, shows peak at the characteristic plasma oscillation energy. The position of the peak marks the energy at which the reflectivity and absorption coefficient falls sharply. Above the plasma energy, the material becomes transparent to the incident photons and the optical features become similar to those of insulators. For $PbTaSe_2$, the plasma peaks are located at 17.6 eV and 16.0 eV for the electric field polarizations along the [100] and [001] directions, respectively.

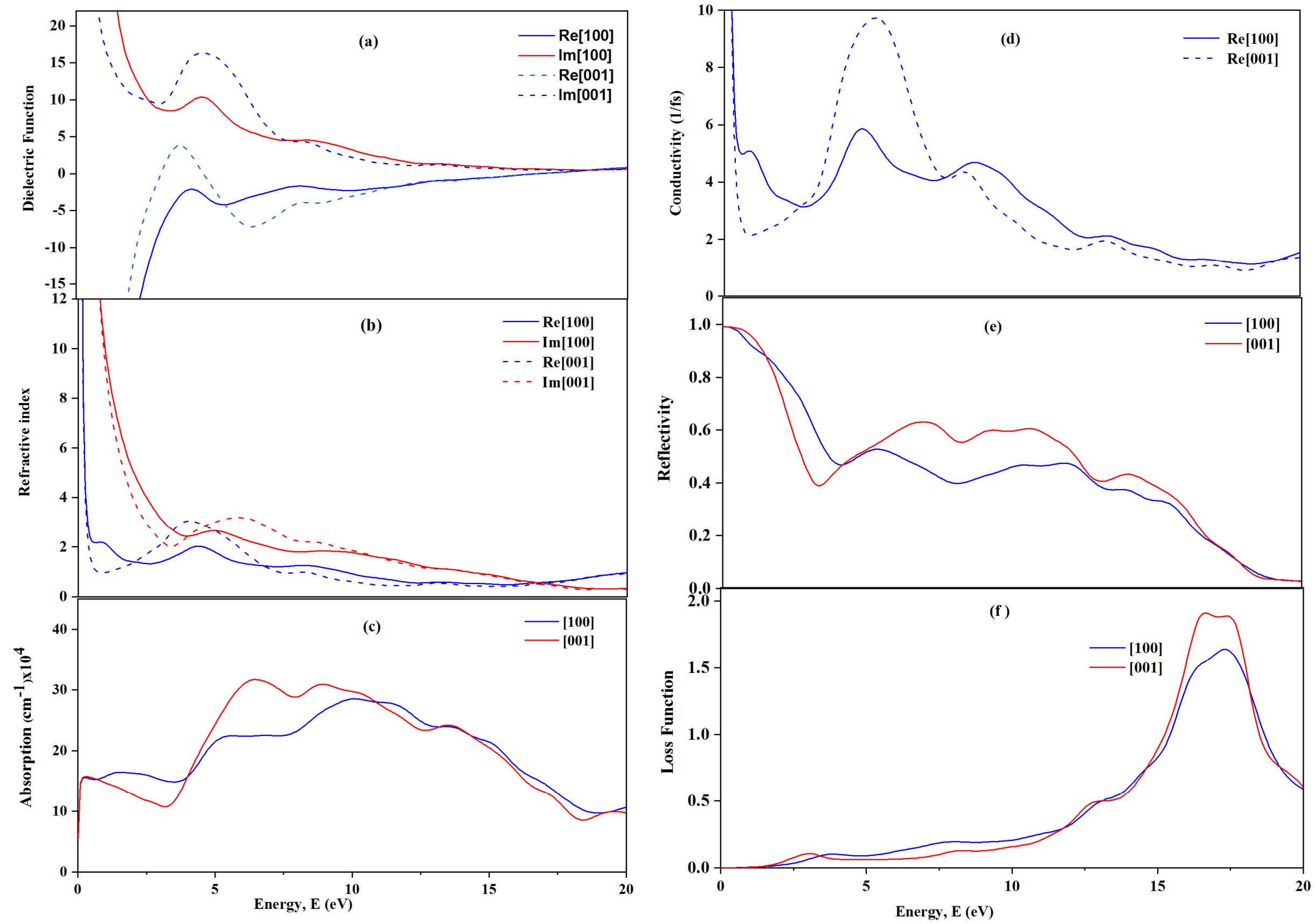

**Figure 7:** The (a) real and imaginary parts of dielectric function [$\varepsilon_1(\omega)$ and $\varepsilon_2(\omega)$], (b) real and imaginary parts of the refractive index [$n(\omega)$ and $K(\omega)$], (c) absorption coefficient [$\alpha(\omega)$], (d) optical conductivity [$\sigma(\omega)$], (e) reflectivity [$R(\omega)$] and (f) loss function [$L(\omega)$] of $PbTaSe_2$ for the [100] and [001] electric field polarizations.

## 4. Conclusions

Using the DFT based first-principles calculations, we have explored a large number of elastic, bonding, lattice dynamic, electronic, thermophysical and optical properties of hexagonal $PbTaSe_2$ topological semimetal. Most of the reported results are novel. The compound is elastically and dynamically stable with ductile features. It is highly machinable and elastically anisotropic. There is significant metallic and ionic bonding in $PbTaSe_2$ with little covalent character. The hardness of $PbTaSe_2$ is moderate. The Debye temperature and phonon thermal conductivity of $PbTaSe_2$ are also low. The electronic band structure shows semimetallic character with bulk topological features. We have found Weyl points in momentum space, where the valence and conduction bands touch each other as found in previous study [9]. The calculated value of the repulsive Coulomb pseudopotential indicates that $PbTaSe_2$ possesses electronic correlations. The Fermi surface has both electron- and hole-like regions. The optical parameters have been explored in detail. The compound under study possesses optical anisotropy. $PbTaSe_2$ is a good absorber of ultraviolet light and reflects visible radiation very effectively. The optical properties also exhibit metallic character and are consistent with the electronic band structure calculations.

**Acknowledgements**

S. H. N. acknowledges the research grant (1151/5/52/RU/Science-07/19-20) from the Faculty of Science, University of Rajshahi, Bangladesh, which partly supported this work.

**Data availability**

The data sets generated and/or analyzed in this study are available from the corresponding author on reasonable request.

**Declaration of interest**

The authors declare that they have no known competing financial interests or personal relationships that could have appeared to influence the work reported in this paper.

## CRediT authorship contribution statement

**A.S.M. Muhasin Reza**: Formal analysis, Methodology, Writing–original draft. **S.H. Naqib**: Supervision, Formal analysis, Conceptualization, Project administration, Writing-review & editing.